\documentclass[3p,times,procedia]{elsarticle}
\usepackage{nupha_ecrc}

\volume{00}

\firstpage{1}

\journalname{Nuclear Physics A}

\jid{nupha}

\jnltitlelogo{Nuclear Physics A}

\usepackage{amssymb}

\usepackage{overpic}

\begin{document}

\begin{frontmatter}

\dochead{XXVIIIth International Conference on Ultrarelativistic Nucleus-Nucleus Collisions\\ (Quark Matter 2019)}

\title{Flow in collisions of light nuclei\tnoteref{label1}}
 \tnotetext[label1]{Talk by WB at parallel session {\em Small systems}, Quark Matter 2019, 4-9 Nov. 2019, Wuhan, China}
 
\author[UJK,IFJPAN]{Wojciech Broniowski}
\ead{Wojciech.Broniowski@ujk.edu.pl}
\author[AGH]{Piotr Bo\.zek}
\ead{Piotr.Bozek@fis.agh.edu.pl}
\author[UJK]{Maciej Rybczy\'nski}
\ead{Maciej.Rybczynski@ujk.edu.pl}
\author[Granada]{Enrique Ruiz Arriola}
\ead{earriola@ugr.es}

\address[UJK]{Institute of Physics, Jan Kochanowski University, 25-406 Kielce, Poland}
\address[IFJPAN]{H. Niewodnicza\'nski Institute of Nuclear Physics, Polish Academy of Sciences, 31-342 Cracow, Poland}
\address[AGH]{AGH University of Science and Technology, Faculty of Physics and Applied Computer Science, 30-059 Cracow, Poland}
\address[Granada]{Departamento de F\'isica At\'omica, Molecular y Nuclear and Instituto Carlos I de F\'{\i}sica Te\'orica y Computacional, \\ Universidad de Granada, E-18071 Granada, Spain}

\begin{abstract}
In this talk we discuss several interesting aspects of collective flow in small systems in reference to planned 
future experiments. First, we bring in the novel possibility of exploring elliptic flow 
with heavy nuclei collisions on polarized deuteron targets, accessible in AFTER@LHC program after Long Shutdown 3. Then we pass
to the Glauber model predictions for the presently considered ${}^{16}{\rm O}-{}^{16}{\rm O}$ program and RHIC and the LHC. 
Finally, we turn to our previous proposal concerning specific 
flow signatures of intrinsic correlations (such as the supposed $\alpha$ clusterization) in the structure of 
light nuclei in collisions with heavy nuclei. In particular, ${}^{12}{\rm C}$--heavy nucleus ultrarelativistic reactions would provide 
insight into the ground-state ${}^{12}{\rm C}$ structure, complementary to the traditional nuclear structure features.
\end{abstract}


\end{frontmatter}

\bigskip \bigskip

\section{Collisions on polarized deuteron targets \label{sec:deut}}

\begin{figure} 
  \hspace*{6.7cm}
  \begin{overpic}[width=0.62\textwidth]{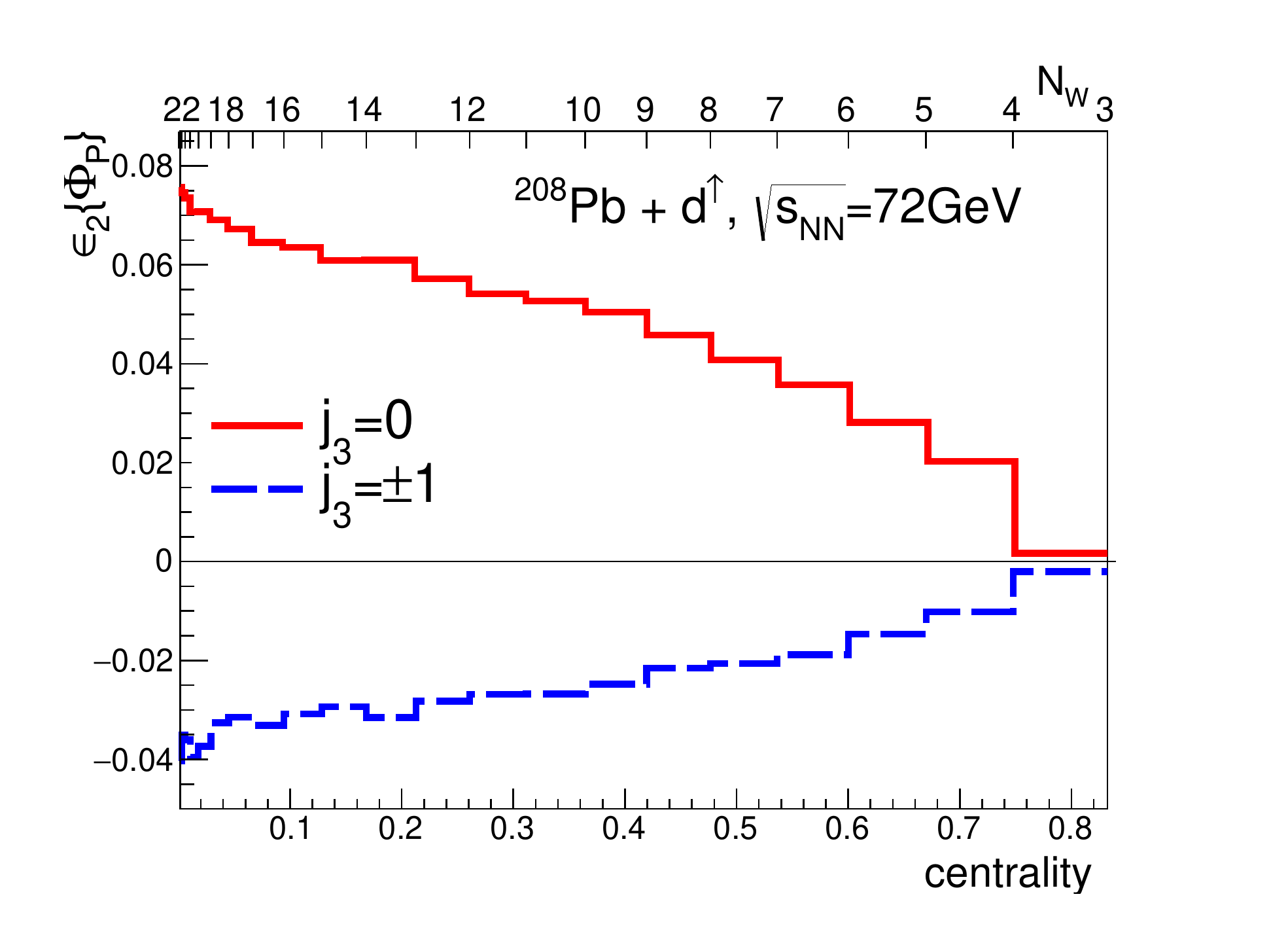}
     \put(-73,25){\includegraphics[width=0.22\textwidth]{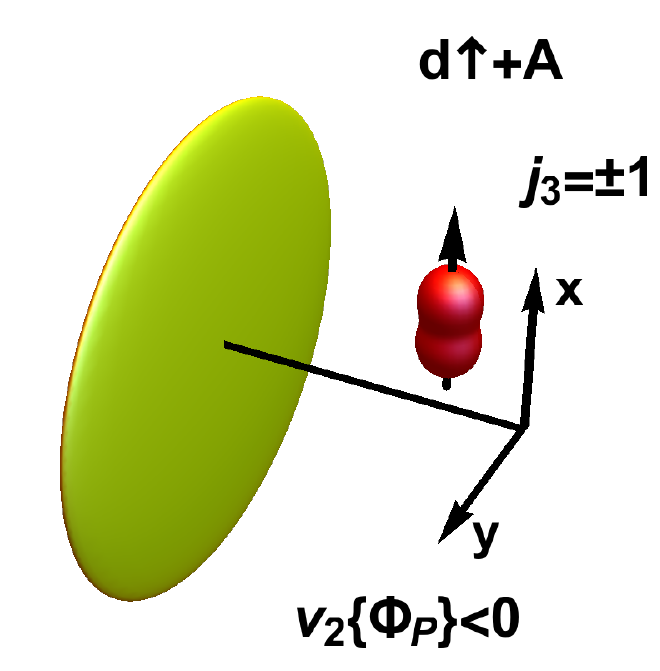} \includegraphics[width=0.22\textwidth]{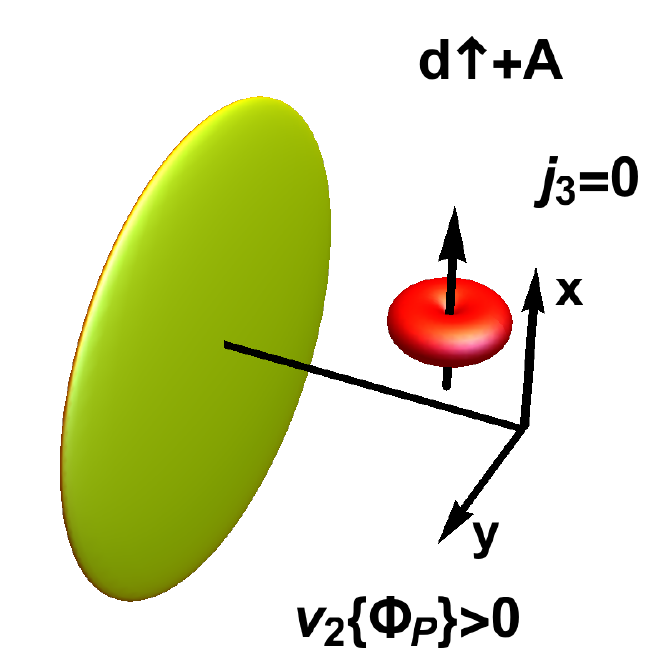}}  
  \end{overpic}
\vspace*{-10mm}  
\caption{\label{fig:d} Left:~A schematic view of the ultra-relativistic collision of a heavy nucleus on the deuteron target
polarized along ($j_3=\pm 1$) and perpendicular ($j_3=0$) to the fixed polarization axis $\{\Phi_P\}$. 
The deformation of the created fireball in the transverse plane reflects the intrinsic deformation of the polarized deuteron. 
The collective {\em shape-flow transmutation} mechanism results in the {\em one body} elliptic flow 
coefficient with respect to the polarization axis, $v_2\{\Phi_P\}$, with the signs as labeled in the figure. 
Right:~Ellipticities of the initial condition in the fireball, evaluated 
with respect to the fixed polarization axis, $\epsilon_2\{\Phi_P\}$, for Pb collisions on a polarized deuteron target at $\sqrt{s_{NN}}=72$~GeV. 
The lower-axis coordinate is the centrality determined from the initial entropy $S$, whereas the 
top-axis coordinate is the corresponding number of the wounded nucleons, $N_W$. (Graphics from~\cite{Broniowski:2019kjo})}
\end{figure}

The deuteron, being a spin $j=1$ nucleus, can be polarized in an external magnetic field $B$. The admixture of the $D$-wave in its 
ground-state wave function 
leads to an intrinsic deformation of the nucleon matter distribution; it is prolate for $j_3=\pm 1$ and 
oblate for for $j_3=0$, where $j_3$ is the projection of spin along the polarization axis $B$ 
(see Fig.~\ref{fig:d}). When such a polarized deuteron target is hit with an ultrarelativistic heavy nucleus, 
the deformation of the formed fireball in the transverse plane 
reflects, to a large degree, the intrinsic deformation of the deuteron. The following collective evolution produces, via the 
shape--flow transmutation mechanism,  
elliptic flow which can be quantified with respect to the (fixed) polarization axis in terms of a {\em one-body} quantity, which
would be straightforward to measure experimentally~\cite{Bozek:2018xzy,Broniowski:2019kjo}.

The planned fixed target AFTER@LHC experiments, in particular SMOG2@LHCb~\cite{Barschel:2014iua,Aaij:2014ida,LHCbCollaboration:2673690,Bursche:2649878}, 
will be able to study collisions of a 2.76A~TeV Pb beam on fixed targets,
with a possibility of using in the future polarized hydrogen and deuterium targets~\cite{Aidala:2019pit}, which can  be 
installed during the LHC Long Shutdown~3 in the years 2023-2025. 
We note that the proposed method requires a measurement of 
a one-body distribution and, with a  very high intensity beam, could be simply 
performed with minimum bias events and without event reconstruction or pile-up corrections. 
Precise estimates, including hydrodynamic simulations, error estimates, etc., are provided in~\cite{Broniowski:2019kjo}.

An analogous effect is present for collisions on other light targets 
with $j\ge 1$,  such as ${}^7$Li, ${}^9$Be, or ${}^{10}$B.  Interestingly, the magnitude of the elliptic flow
can be estimated from their known mean square radii and  quadrupole moments, and is sizable, even larger that for the case 
of the deuteron. 
The estimate for the elliptic flow coefficient 
evaluated with respect to the polarization axis is~\cite{Broniowski:2019kjo}
\begin{eqnarray}
v_2\{\Phi_P\} \simeq - k \frac{3 Q_2}{4Z( \langle r^2 \rangle+ \frac{3}{2}\langle b^2 \rangle)} \frac{3j_3^2-j(j+1)}{j(2j-1)}, \nonumber
\end{eqnarray}
where $k\sim 0.1$ is the hydrodynamic response coefficient, $Q_2$ is the quadrupole moment, $Z$ is the atomic number, and $\langle r^2 \rangle$ 
is that mean squared charge radius of the light nucleus. The quantity $\langle b^2 \rangle \sim 1~{\rm fm}^2$ is the average impact parameter squared in 
inelastic $NN$ collisions.  The formula holds for 
perfect polarization, sufficiently central collisions, and $j \ge 1$.

If the effect of the elliptic flow in polarized heavy--light collisions is indeed confirmed, it would corroborate the scenario of the  
late-stage generation of collectivity.
Other interesting opportunities emerging from such collisions involve studies of hard probes as well as femtoscopic 
correlations, with appropriate measures defined with respect to the polarization axis.

\section{${}^{16}{\rm O}-{}^{16}{\rm O}$ collisions \label{sec:OO}}

\begin{figure}
\includegraphics[angle=0,width=0.43 \textwidth]{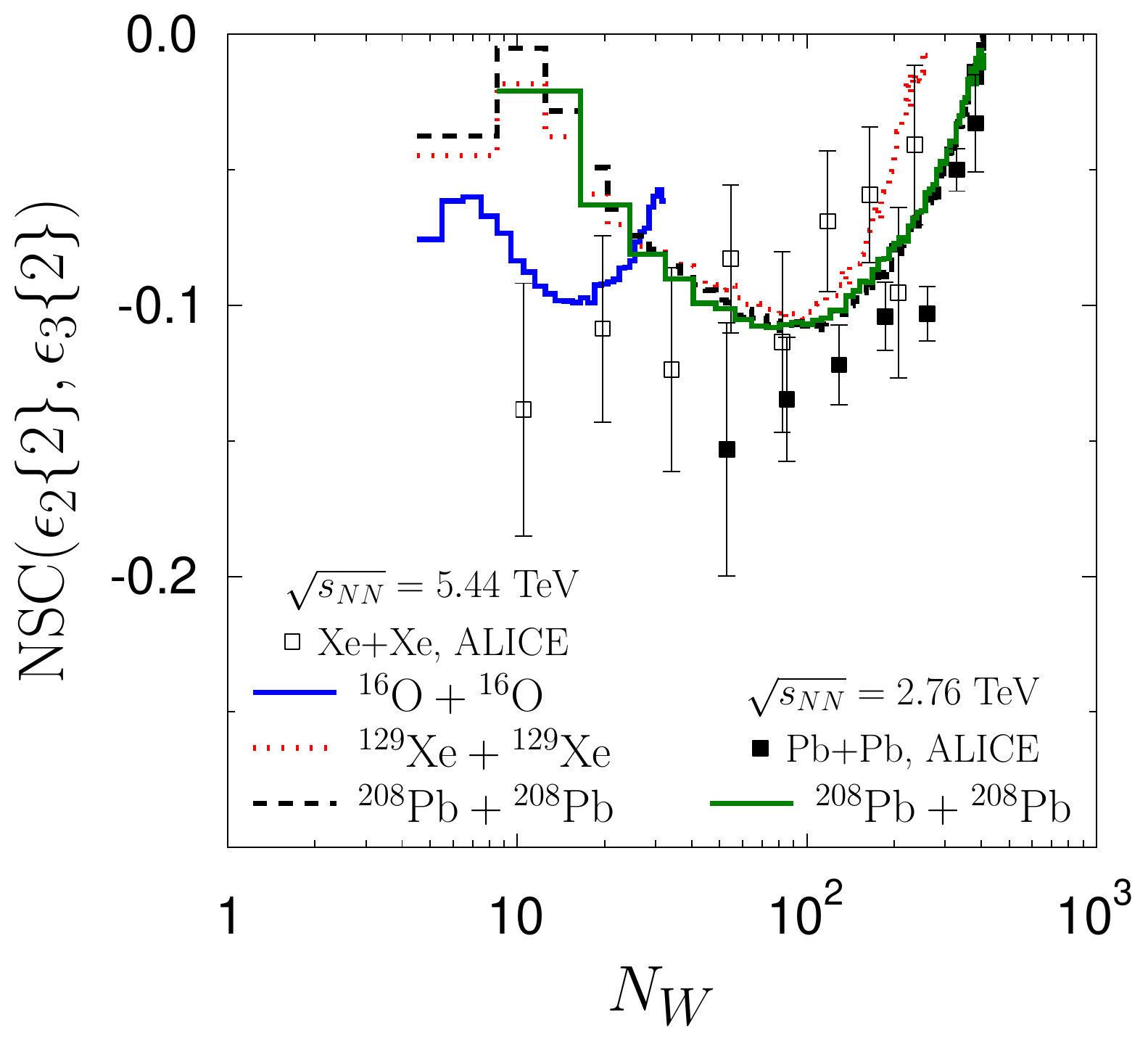} \hfill
\includegraphics[angle=0,width=0.43 \textwidth]{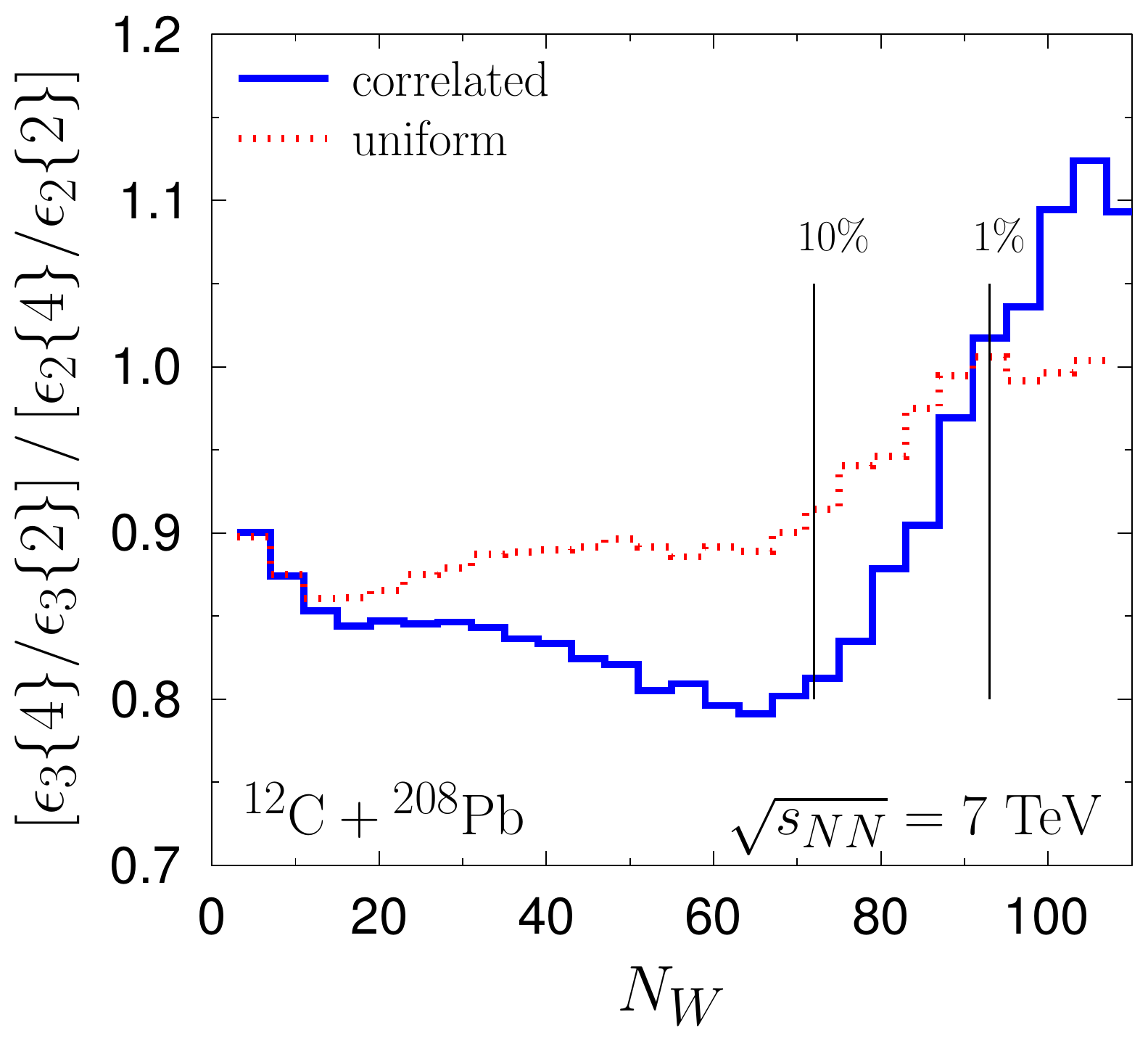}
\vspace{-2mm}
\caption{\label{fig:OO} Left:~{\tt GLISSANDO}~\cite{Bozek:2019wyr} results for 
symmetric cumulants for various collisions systems indicated in the legend,
plotted as a function of $N_{\rm w}$. Large systems are compared to the available data. The solid line represents predictions for 
${}^{16}{\rm O}-{}^{16}{\rm O}$ collisions. 
Right:~Double eccentricity ratio for ${}^{12}{\rm C}-{}^{208}{\rm Pb}$ collisions from {\tt GLISSANDO} 
for the correlated (clustered) and uniform ${}^{12}{\rm C}$ structure,
plotted as a function of $N_{\rm w}$. The observed non-monotonicity of the solid line is a hallmark of clusterization. 
(Graphics in the left panel from~\cite{Rybczynski:2019adt})}
\end{figure}

Proposals to study collisions with ${}^{16}{\rm O}$ beams at the LHC~\cite{Citron:2018lsq}
and at RHIC~\cite{starO16} are presently under serious consideration. In this regard we have 
carried out an analysis of the initial state in ${}^{16}{\rm O}$-${}^{16}{\rm O}$ in the Monte Carlo Glauber approach~\cite{Rybczynski:2019adt}. Similar results 
in other models were presented earlier in~\cite{Sievert:2019zjr,Huang:2019tgz}. The results can be summarized with a 
statement that they are qualitatively similar to the effects find in heavier systems, with a natural scaling towards the smaller 
number of participants. A typical result is shown in the left panel of Fig.~\ref{fig:OO}, where we plot  the normalized symmetric cumulant,
measuring correlations between the elliptic and triangular deformations. We note a similar behavior in ${}^{16}{\rm O}$-${}^{16}{\rm O}$ 
as in Pb-Pb or Xe-Xe, albeit, as expected, moved to 
lower values of the number of wounded nucleon, $N_{\rm W}$. Further results are given in~\cite{Rybczynski:2019adt}.

\section{Signature of intrinsic correlations \label{sec:alpha}}

Over the past few years we have been pursuing the idea~\cite{Broniowski:2013dia,Arriola:2014lxa,Bozek:2014cva,Rybczynski:2017nrx} 
that clustering in light nuclei may produce specific signals in harmonic flow correlations in reactions with heavy nuclei.  
The issue is fundamentally interesting, as in collisions at highest available energies the 
reaction time is so short that the existing nucleon cluster
structures (such as the $\alpha$ particles present in the ground state) are effectively 
frozen when the nuclear wave function collapses.  
As a result, the shape of the formed fireball in the transverse plane is, event-by-event, revealing the information on the lowest-energy
ground-state of the light nucleus colliding with a ``wall'' of a heavy nucleus. 

Most promising signals for the cluster effects would originate from $^{12}$C - heavy nucleus collisions, 
since the $^{12}$C nucleus is believed to have a prominent cluster structure, with three $\alpha$ particles placed in 
corners of an equilateral triangle.
In our {\tt GLISSANDO}~\cite{Bozek:2019wyr} simulations we have used configurations from state-of-the-art cluster 
Variational Monte Carlo simulations~\cite{Lonardoni:2017egu}, as supplied 
in files in~\cite{Loizides:2014vua}. Similar studies were reported in~\cite{Lim:2018huo}. 
We stress that the used dynamically generated distributions contain realistic nuclear correlations.

An example of a measure sensitive to the cluster correlations in the light nucleus 
is the double ratio of triangular and elliptic eccentricities evaluated with cumulants with 2 and 4 particles,  
$[\epsilon_3\{4\}/\epsilon_3\{2\}]/[\epsilon_2\{4\}/\epsilon_2\{2\}]$. It is plotted in
the right panel of Fig.~\ref{fig:OO} as a function of the number of wounded nucleons. We note a characteristic non-monotonic behavior for the case where 
the $^{12}$C nucleus is clustered (solid line), as opposed to the case of a uniform distribution (dotted line). 
An analogous behavior is expected for the corresponding double ratio of the elliptic and triangular flow coefficients.  
More details can be found in~\cite{Broniowski:2013dia,Arriola:2014lxa,Bozek:2014cva,Rybczynski:2017nrx}.

\bigskip

To summarize, various intriguing and hitherto unexplored physical effects may be explored in future heavy-light ultrarelativistic nuclear collisions, 
which may further probe the issue of the origin of collectivity in the fireball dynamics and its limits, as well as investigate the nuclear structure 
correlation effects in the initial state. The observation of the discussed effects would confirm the scenario of the 
late stage generation of collective flow in small systems, which, while plausible, needs independent and unique tests.

\bigskip \bigskip \bigskip
Project supported by the Polish National Science Centre grants 2015/19/B/ST2/00937 (WB), 2018/29/B/\-ST2/00244 (PB), 2016/23/B/ST2/00692 (MR), and
by the Spanish Ministerio de Economia y Competitividad and European FEDER funds (grant FIS2017-85053-C2-1-P) and 
Junta de Andaluc\'{\i}a grant FQM-225 (ERA).

\bibliographystyle{elsarticle-num}
\bibliography{hydr}

\end{document}